\documentclass[%
 reprint,
 amsmath,amssymb,
 aps,
]{revtex4-1}

\usepackage{mathrsfs}
\usepackage{amsmath}
\usepackage{indentfirst}
\usepackage{amssymb}
\usepackage{color}
\usepackage{amsfonts}
\usepackage[framemethod=TikZ]{mdframed}
\usepackage{mathtools}

\usepackage[retainorgcmds]{IEEEtrantools}
\usepackage[colorlinks=true,linkcolor=blue,urlcolor=blue,citecolor=blue,pdfusetitle]{hyperref}
\usepackage{times,txfonts}
\usepackage{graphicx}
\usepackage{dcolumn}
\usepackage{bm}

\usepackage[caption=false]{subfig}

\usepackage{xcolor}

\usepackage{hyperref}
\newcommand{\tr}{\text{tr}}
\usepackage{comment}

\newcommand\mydots{\hbox to 1em{.\hss.\hss.}}
\newcommand{\GTL}[1]{\textcolor{orange}{GTL: #1}}

\begin{document}


\title{Stroboscopic two-stroke quantum heat engines}%

\author{Otavio A. D. Molitor}
\email{otavio.molitor@usp.br}
 \affiliation{%
Instituto de Física da Universidade de São Paulo, 05314-970 São Paulo, Brazil
}%
\author{Gabriel T. Landi}%
\email{gtlandi@if.usp.br}
\affiliation{%
Instituto de Física da Universidade de São Paulo, 05314-970 São Paulo, Brazil
}%

\date{\today}

\begin{abstract}

The formulation of models describing quantum versions of heat engines plays an important role in the quest toward establishing the laws of thermodynamics in the quantum regime.  Of particular importance is the description of stroke-based engines which can operate at finite-time.
In this paper we put forth a framework for describing stroboscopic, two-stroke engines, in generic quantum chains. 
The framework is a generalization of the so-called SWAP engine and is based on a collisional model, which alternates between pure heat and pure work strokes.
{\color{black} The transient evolution towards a limit-cycle is also fully accounted for. Moreover, we show that once the limit-cycle has been reached, the energy of the internal sites of the chain no longer changes, with the heat currents being associated exclusively to the boundary sites.}
Using a combination of analytical and numerical methods, we 
show that this type of engine offers multiple ways of optimizing the output power, without affecting the efficiency.
{\color{black}Finally, we also show that there exists an entire class of models, characterized by a specific type of inter-chain interaction, which always operate at Otto efficiency, irrespective of the operating conditions of the reservoirs.}


\end{abstract}

\maketitle

\section{\label{sec:intro}Introduction}

\begin{figure*}
    \centering
    \includegraphics[scale=0.46]{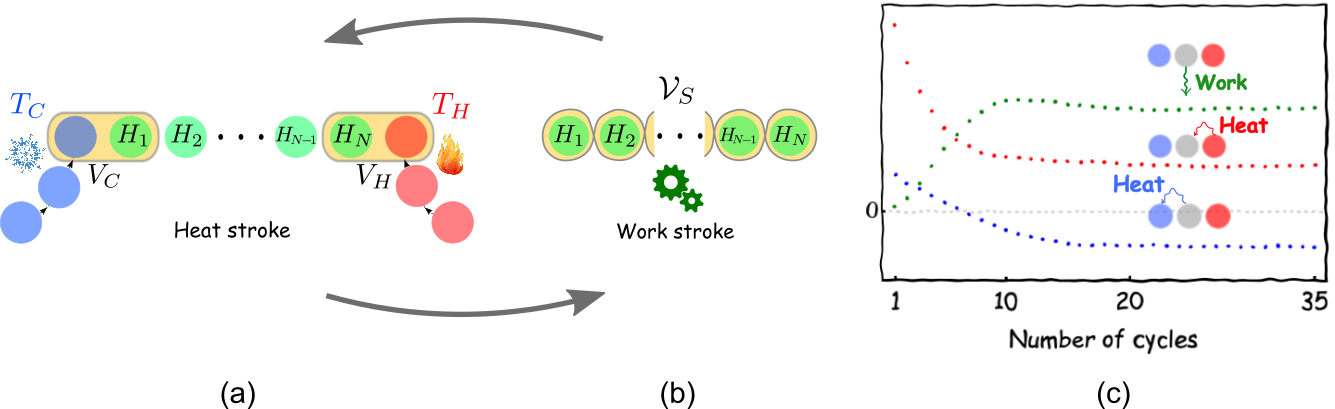}
    \caption{The quantum heat engine (QHE) model is composed of two processes: (a) Heat stroke ($q$) and (b) Work stroke ($w$). During the heat stroke the internal interactions of the quantum chain are turned off while the boundary subsystems 1 and $N$ interact with a cold bath at temperature $T_C$ and a hot bath at temperature $T_H$, respectively. The work stroke, in its turn, consists of a unitary evolution of the  chain as a whole, by means of the interaction $\mathcal{V}_S$, and with the baths disconnected from the chain. The strokes are operated sequentially in a cyclic way, as shown by the arrows around (a) and (b). The result is a stroboscopic evolution of the state of the chain, which results in a discrete-time evolution of all thermodynamic quantities, as illustrated in image (c).}
    \label{fig:panel}
\end{figure*}

One of the cornerstones in the theoretical formulation of quantum thermodynamics~\cite{Binder2018} is the development of Quantum Heat Engines (QHEs) containing quantum systems as the working fluid~\cite{Geva1992,Kosloff2002,Quan2007,Campisi2015,Kosloff2017,Scopa2018,Chen2019a,Linden2010,Kosloff2014,Muller2018,DeChiara2018,Clivaz2019,Mitchison2019,Insinga2018,Josefsson2020,Gherardini2020,Rezek,Kim2011,PhysRevE.87.012140}. 
The goal is to extend the notion of thermodynamic cycles to the quantum regime, with the aim of not only designing ultrasmall engines and optimizing them, but also understanding, at a more fundamental level, the limits of energy conversion in the quantum regime. 
As in classical thermodynamics, QHEs may be classified as either operating in continuous-time~\cite{Linden2010,Kosloff2014,Muller2018,DeChiara2018,Clivaz2019,Mitchison2019} or being stroke-based~\cite{Geva1992,Kosloff2002,Quan2007,Campisi2015,Kosloff2017,Scopa2018,Chen2019a}. 
Continuous engines, such as thermoelectrics or  masers~\cite{Scovil1959,Kosloff2014}, operate autonomously and extract work in the form of steady currents (e.g. chemical work in the case of thermoelectrics).
Stroke-based engines, on the other hand, are based on a series of alternating steps that form the thermodynamic cycle.
Work is performed by changing the system Hamiltonian while heat is exchanged by coupling the system to alternating baths.
Time is thus explicitly treated as a variable, which can be used to optimize the output power.


The theoretical modeling of quantum heat engines, however, quickly stumbles on serious mathematical complications. 
The coupling to heat baths is usually done using master equations or quantum operations, which often rely on several approximations. 
And while these may only have a mild effect on the dynamics, they may profoundly affect the \emph{thermo}dynamics. 
The reason is ultimately related to energy conservation; that is, in making sure that all energy sources and sinks are appropriately taken into account and properly identified as either heat or work.
While this is usually easy in classical systems, 
in the quantum realm it becomes extremely delicate. 

In most thermodynamic cycles, work is performed while the system is in contact with one or more baths. 
From a modeling point of view, this requires the derivation of time-dependent master equations, which can be done using e.g. Floquet theory, but is generally quite involved. 
For this reason, most studies have focused on the Otto cycle, where heat and work strokes are clearly separated: During part of the dynamics, the system is coupled to a bath and allowed to relax with a fixed Hamiltonian, while in others the system is isolated and the Hamiltonian is driven externally. 
Even in this case, however, problems may arise since the act of coupling and uncoupling the system from the baths can have an associated work cost~\cite{DeChiara2018,Barra2015,Pereira2018}. 
This is related to the size of the system-bath interaction, when compared with the typical energy of the system, something which can be significant in quantum systems. 

The importance of controlling all the energy sources for a consistent treatment of the thermodynamics of quantum systems is well addressed in resource theories of thermodynamics~\cite{Sparaciari2017,Brandao2013,Gour2015}. In this formulation of thermodynamics, the quantities are treated using tools of quantum information theory, such as global energy-preserving unitaries, known as Thermal Operations, that guarantee full control over the system and the fulfillment of the 1$^\text{st}$ law of thermodynamics at all times~\cite{Sparaciari2017}. In this framework, continuous-time QHEs are naturally implemented, since all energy sources are embedded into the system and there is no need of an external agent to operate it, which would impose a difficulty in keeping track of all energy sources. However, when it comes to QHEs that are externally operated through strokes, the accountability of every single source of energy becomes a challenge, if one considers master equations. 

To shed light on this issue, it is  essential to consider models where all energy sources are properly taken into account.
With this in mind, we put forth in this paper a general framework for dealing with stroke-based QHEs operating with only two strokes.
The framework is based on the idea of collisional models~\cite{Scarani2002,Strasberg2019a,Landi2014,Giovannetti2012,Pereira2018,Ziman2002,Rodrigues2019a}, where the reservoirs are modeled by identical and independently prepared (iid) units (henceforth called ancillas), which interact with the system one at a time. 
The basic idea is illustrated in Fig.\ref{fig:panel}. 
We consider a quantum chain with $N$ sites, each with local Hamiltonian $H_i$ and interacting according to some interaction Hamiltonian $V$ which, for concreteness, we take to be nearest neighbor interactions; that is $\mathcal{V}_S = \sum_i V_{i,i+1}$ (although all results also hold for longer ranged interactions). 
The chain is also connected to two baths at each end (the generalization to more baths is also straightforward). 
Each bath is composed by ancillas with Hamiltonians $H_C$ and $H_H$ and prepared in thermal states $\rho_{x} = e^{-H_x/T_x}/Z_x$ (with $x = C,H$) at different temperatures $T_C$ and $T_H$. 
Non-thermal reservoirs can also in principle  be implemented, using for instance the results of~\cite{Rodrigues2019a}. 

The engine operates in two strokes.
The first is the heat stroke ($q$), where the internal interaction $\mathcal{V}_S$ is turned off and the system is allowed to interact with $C$ and $H$ (Fig.~\ref{fig:panel}(a)). 
The thermodynamic analysis of this kind of process is by now well established~\cite{DeChiara2018,Barra2015,Pereira2018} and any potential work sources stemming from turning the $CS\!H$ interaction on and off, can be properly taken into account. 
In the second stroke ($w$), the system is completely isolated and the interaction $\mathcal{V}_S$ is turned on for a certain amount of time. 
This allows currents to flow through the chain, which is associated with a certain amount of work (Fig.~\ref{fig:panel}(b)).
This scenario can be viewed as a generalization of the so-called SWAP engine~\cite{Allahverdyan2010,Campisi2014,Uzdin2014,Campisi2015,Guarnieri2019}, in which the system is composed of only two qubits and the interactions are partial SWAPs.
Here the number of sites in the chain is arbitrary, as well as the form of the interactions.

Our construction is particularly suited for modeling finite-time effects. 
The typical dynamics of heat and work is illustrated in Fig.~\ref{fig:panel}(c). 
As soon as the engine is turned on, all quantities will undergo a transient (stroboscopic) dynamics. 
After a sufficiently large number of cycles, however, they converge to a limit-cycle, where the engine's operation becomes periodic (the stroboscopic analog of a non-equilibrium steady-state).

The paper is organized as follows. 
The QHE model is presented in Sec.\ref{sec:2strokemodel}, where we lay the basic expressions for all relevant thermodynamic quantities. 
In particular, we also show that, depending on the type of interaction $\mathcal{V}_S$, the efficiency of the QHE may have a universal value, independent of the operating conditions.
In Sec.~\ref{sec:applications}, the framework is then applied to two concrete examples: a two-qubit QHE (Sec.~\ref{sec:analytic}) and a  spin chain with $N$ sites (Sec.~\ref{sec:Nspins}). 
The former, in particular, is treated analytically, by casting the problem as a set of difference equations for some relevant system operators.
In both cases, we explore how the parameter space affects the output power, as well as the number of cycles needed to attain the limit-cycle regime. Finally, concluding remarks are presented in Sec.\ref{sec:conclusion}.

\section{\label{sec:2strokemodel}Two-stroke quantum heat engine}

In this section we present a detailed description of the proposed two-stroke engine. 
We start by separately describing the heat and work strokes, which are then sewn together to yield the complete stroboscopic dynamics.
Here and henceforth,  all quantities are expressed in  units of $k_B=\hbar=1$.

\vspace{-0.5cm}

\subsection{\label{sec:heatstroke}Heat stroke}

The heat stroke is depicted in Fig.~\ref{fig:panel}(a).
The working fluid (henceforth referred to as ``the system'') is composed of $N$ sites, each with dimension $d_i$ and local Hamiltonians $H_i$. The $N$ sites are initially prepared in an arbitrary state $\rho_S$, which need not be a product. 
During the heat stroke, the sites do not interact in any way (although the state $\rho_S$ may very well be non-local). 
Each heat stroke is characterized by the interaction with two baths, $C$ and $H$, at the boundaries. 
The baths are described by iid ancillas, each with local Hamiltonian $H_x$ and prepared in thermal states $\rho_{x} = e^{-H_x/T_x}/Z_x$ (with $x = C,H$) at different temperatures $T_C$ and $T_H$ (for concreteness, we set $T_C < T_H$).
The interaction Hamiltonian $V_C$ of the left bath has support only between $C$ and subsystem $S_1$, while $V_H$ has support on $S_N$ and $H$. 
This interaction is characterized by the global unitary 
\begin{equation}
    \tilde{\rho}_S = \text{tr}_{CH} \big \{ U_q (\rho_{C}\hspace{0.5pt} \rho_S\hspace{0.5pt} \rho_{H}) U_q^\dagger \big \} :=\mathcal{E}_q(\rho_S),
    \label{eq:mapQ}
\end{equation}
where $U_q = e^{-i H_q \tau_q}$, $\tau_q$ is the  duration of the stroke and  
$H_q= \sum_{i=1}^{N} H_{i}+H_{C}+H_H+V_C+V_H$.

The main advantage of the collisional model approach is the ability to properly account for all changes in energy in both system and baths. 
We \emph{define} heat as minus the change in energy of the ancillas, 
\begin{equation}
    \mathcal{Q}_x := -\tr\big\{ H_x (\tilde{\rho}_x - \rho_x) \big\}, 
    \qquad x = C,H,
    \label{heat_A}
\end{equation}
where $\tilde{\rho}_x$ is the reduced state of $C$ or $H$ after the map~\eqref{eq:mapQ}. 
In general, however, this will not equal the change in energy of the system. 
The reason is that turning the interactions $V_x$ on and off will, in general, have an associated energy cost, called the ``\emph{on/off work}''~\cite{DeChiara2018}. 
In fact, energy conservation for each individual stroke implies that 
\begin{IEEEeqnarray}{rCl}
\label{Wc_on_off}
    W_C^\text{on/off} &:=& \mathcal{Q}_C + \tr\big\{H_1 (\tilde{\rho}_S - \rho_S) \big\} = - \Delta V_C,\\[0.2cm]
    W_H^\text{on/off} &:=& \mathcal{Q}_H + \tr\big\{H_N (\tilde{\rho}_S - \rho_S) \big\} = - \Delta V_H,    
\label{Wh_on_ff}
\end{IEEEeqnarray}
where $\Delta V_x = \tr\big\{V_x (\tilde{\rho}_{CS\!H} - \rho_C\rho_S\rho_H)\big\}$.
The on/off work is thus associated with energy that stays ``trapped'' in the interactions $V_x$. 

The condition required for the on/off work to be zero is called strict energy conservation and reads
\begin{equation}
    [V_C,H_{1}+H_{C}]=[V_H,H_{N}+H_{H}]=0.
    \label{eq:localbalance}
\end{equation}
{\color{black}To make the paper more self-contained, we provide a simple proof of this in appendix~\ref{app:SEC}.}
In the language of resource theories, Eq.~\eqref{eq:localbalance} means that the map~\eqref{eq:mapQ} is a combination of two thermal operations~\cite{Horodecki2013,Brandao2013,Brandao2015}, one acting on site 1 and the other on site $N$. 
When~\eqref{eq:localbalance} is satisfied, all energy leaving the system must enter the baths and vice-versa. 
As a consequence, the heat~\eqref{heat_A} may be equivalently defined as 
\begin{equation}
    \mathcal{Q}_C := \tr\big\{ H_1 (\tilde{\rho}_S - \rho_S) \big\}, 
    \qquad
    \mathcal{Q}_H := \tr\big\{ H_N (\tilde{\rho}_S - \rho_S) \big\}, 
    \label{heat}
\end{equation}
which can now be computed solely from knowledge of the reduced state of the system.

A popular choice of interactions are those which have the form
\begin{equation}\label{VC_eigenop}
    V_C = \sum_{k} g_k\big ( L_{k}^\dagger A_{k} + L_{k} A_{k}^\dagger \big ),
\end{equation}
where $L_k$ is an operator acting only on subsystem $1$ and $A_k$ are operators acting on $C$. 
A similar definition holds for the interaction $V_H$ between $H$ and site $N$.
The condition~\eqref{eq:localbalance} can be fulfilled in this case whenever the $\{L_k\}$ and $\{A_k\}$ are eigenoperators of $H_1$ and $H_C$; 
that is, if they satisfy 
$[H_{S},L_{k}]=-\omega_{k}L_{k}$ and $[H_{E},A_{k}]=-\omega_{k}A_{k}$, 
for the \emph{same} set of frequencies $\{\omega_k\}$.
In this case one usually says that $C$ and $1$ are \emph{resonant}, meaning that all energy that leaves one enters the other. 

We will not assume that the interaction is necessarily of the form~\eqref{VC_eigenop}, but we will from now on assume that strict energy conservation~\eqref{eq:localbalance} is satisfied. 
As a consequence $W_x^\text{on/off} \equiv 0$ and therefore the change in energy of the system  during the heat stroke can unambiguously be associated to heat flowing to and from the reservoirs. 
Finally, we also mention that in the end of the heat stroke, the reservoir ancillas are discarded and never participate again in the dynamics. 
This is another convenience of collisional models: since they are subsequently discarded, one can make any desired measurements in the ancillas, without having to worry about a possible measurement backaction~\cite{Santos2020}.

\subsection{\label{sec:workstroke}Work stroke}

In a similar fashion, we now characterize the work stroke (Fig.~\ref{fig:panel}(b)). 
The system is now  isolated from the rest of the world and its subsystems are put to interact by means of an interaction Hamiltonian $\mathcal{V}_S=\sum_{i}V_{i,i+1}$, which is turned on only during the work stroke.
The system will therefore evolve according to 
\begin{equation}
    \rho_S' = U_w \tilde{\rho}_S U_w^\dagger=\mathcal{E}_w(\tilde{\rho}_S),
    \label{eq:mapw}
\end{equation}
where $U_w=e^{-iH_w\tau_w}$, $H_w = \sum_{i=1}^N H_{i} + \mathcal{V}_S$ and $\tau_w$ is the duration of the work stroke. 

During this stroke, by turning on $\mathcal{V}_S$, currents are allowed to flow through the system (which will eventually flow to the reservoirs in the next stroke).
The work cost associated to this is simply the on/off work of turning $\mathcal{V}_S$ on and off; viz., 
\begin{equation}
    \mathcal{W} = -\tr\bigg\{ \big(\sum_i H_i\big) (\rho_S' - \tilde{\rho}_S) \bigg\} = \tr \big\{ \mathcal{V}_S (\rho_S' - \tilde{\rho}_S)\big\}. 
    \label{work}
\end{equation}
Work is defined as positive when energy leaves the system (i.e. work is \emph{extracted}), while the heats $\mathcal{Q}_x$ in Eq.~\eqref{heat} are positive when energy enters the system. 

One can also offer the following alternative justification for Eq.~\eqref{work}~\cite{DeChiara2018,Pereira2018}. 
Strictly speaking, $\mathcal{W}$ is associated with turning on and off the interaction $\mathcal{V}_S$. 
The system Hamiltonian should thus be taken to be time-dependent, of the form 
\[
H_S(t) = \sum_i H_i + \lambda(t) \mathcal{V}_S,
\]
where $\lambda(t)$ is a boxcar function, taking the value 1 in a window of time $\tau_w$. 
Focusing only on a single stroke, the work can then be defined using the standard statistical mechanics expression 
\[
    \mathcal{W} = -\int\limits_{-\infty}^\infty \left\langle \frac{\partial H_S(t)}{\partial t} \right\rangle dt. 
\]
Since the only time-dependence is in the boxcar $\lambda(t)$ (whose derivative is a pair of $\delta$ functions), one then readily finds that $\mathcal{W}$ is given precisely by Eq.~\eqref{work}.

\subsection{\label{sec:strob}Stroboscopic dynamics}

The result of sewing together the two strokes is a cycle with period $\tau=\tau_q+\tau_w$. 
We let $\rho_S^n$ denote the state of the system after the $n$-th cycle. Combining Eqs.~\eqref{eq:mapQ} and \eqref{eq:mapw} one then finds that $\rho_S^n$ will evolve stroboscopically according to 
\begin{IEEEeqnarray}{rCl}
\label{stroke1}
    \tilde{\rho}_S^{n} &=& \mathcal{E}_q(\rho_S^n), \\[0.2cm]
    \rho_S^{n+1} &=& \mathcal{E}_w(\tilde{\rho}_S^n) 
    = \mathcal{E}_w \circ \mathcal{E}_q (\rho_S^n),
\label{stroke2}
\end{IEEEeqnarray}
for $n \in \mathbb{Z}$. 
The notation $\tilde{\rho}_S^n$ is used to denote the intermediate state, in between the two strokes.

The heat and work in each stroke will be denoted by $\mathcal{Q}_x^n$ and $\mathcal{W}^n$.
They are readily computed from Eqs.~\eqref{heat} and \eqref{work} respectively. 
The first law for the system thus becomes
\begin{equation}\label{1st_law}
    \Delta E_n = \mathcal{Q}_C^n + \mathcal{Q}_H^n - \mathcal{W}^n, 
\end{equation}
where $\Delta E_n = \tr\big\{ (\sum_i H_i) (\rho_S^{n+1} - \rho_S^n)\big\}$ is the change in energy of the system during stroke $n$. 
Since the energy is a function of state, $\Delta E_n$ can simply be written as the difference between the average energies at each stroke; the same, of course, is not true for $\mathcal{Q}_x^n$ and $\mathcal{W}^n$.

Similarly, one may also write down the 2nd law. 
Entropy is only produced during the heat stroke, so that the 2nd law can be written as~\cite{Strasberg2017} 
\begin{equation}\label{2nd_law}
    \Sigma^n = S(\rho_S^{n+1}) - S(\rho_S^n) - \frac{\mathcal{Q}_C^n}{T_C} - \frac{\mathcal{Q}_H^n}{T_H} \geq 0,    
\end{equation}
where $S(\rho)=-\text{tr}\{\rho \text{ln}\rho \}$ is the von Neumann entropy.
The positivity of $\Sigma^n$ can be readily proven, for instance, by writing it in terms of the mutual information developed between system and ancilla~\cite{Esposito2010a,Timpanaro2019a}. 
In this sense, it is also worth mentioning that this result holds even in the presence of on/off work in the heat stroke, provided $\mathcal{Q}_x$ is associated with the change in energy of the ancillas~\cite{Timpanaro2019a}. 


\subsection{\label{sec:limit_cycle}Limit cycle}

\begin{figure}[t]
    \centering
    \includegraphics[width=0.3\textwidth]{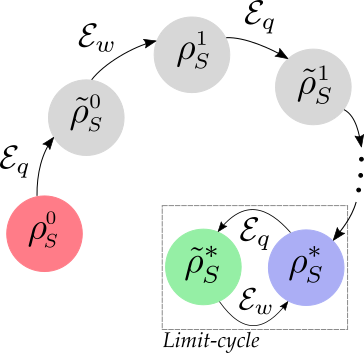}
    \caption{Stroboscopic evolution of the working fluid from its initial state $\rho_S^0$ until it reaches a limit-cycle of two nonequilibrium steady states $\rho_S^*$ and $\tilde{\rho}_S^*$.}
    \label{fig:convergencev3}
\end{figure}

Repeated application of Eq.~\eqref{stroke2} will eventually take the system towards a limit cycle $\rho_S^*$, which is the solution of
\begin{equation}
    \rho_S^* = \mathcal{E}_w \circ \mathcal{E}_q (\rho_S^*).
\end{equation}
The limit cycle is the stroboscopic analog of a non-equilibrium steady-state. 
Crucially, $\rho_S^*$ is a fixed point only of the joint map $\mathcal{E}_w \circ \mathcal{E}_q$, not the individual ones. 
In the limit cycle the system will therefore keep alternating between $\rho_S^*$ and  $\tilde{\rho}_S^* = \mathcal{E}_q(\rho_S^*)$, as depicted in Fig.~\ref{fig:convergencev3}.

In the limit cycle, the first law~\eqref{1st_law} simplifies to 
\begin{equation}\label{1st_law_LC}
    \mathcal{W}^* = \mathcal{Q}_C^* + \mathcal{Q}_H^*,
\end{equation}
meaning that the total heat flux during the heat stroke is converted into a net work at the work stroke. 
Similarly, the second law~\eqref{2nd_law} becomes
\begin{equation}\label{Sigma_LC}
    \Sigma^* = - \frac{\mathcal{Q}_C^*}{T_C} - \frac{\mathcal{Q}_H^*}{T_H}.
\end{equation}
In the standard thermodynamic scenario, the two terms on the RHS are associated with a flow of entropy to each side. 
Thus, in the limit cycle, all entropy produced in the process flows to the environment (because the entropy of the system itself no longer changes). 

A special feature of the limit-cycle in two-stroke engines is that, as illustrated in Fig.~\ref{fig:convergencev3}, the state of the system bounces back and forth between only two states $\rho_S^*$ and $\tilde{\rho}_S^*$. 
The expressions for the heat and work, Eqs.~\eqref{heat} and \eqref{work} thus simplify to 
\begin{IEEEeqnarray}{rCl}
\label{QC_limit}
    \mathcal{Q}_C^* &=& \tr\big\{ H_1 (\tilde{\rho}_S^* - \rho_S^*) \big\}, \\[0.2cm]
\label{QH_limit}    
    \mathcal{Q}_H^* &=& \tr\big\{ H_N (\tilde{\rho}_S^* - \rho_S^*) \big\}, \\[0.2cm]
\label{W_limit}    
    \mathcal{W}^* &=& -\tr\bigg\{ \Big(\sum_i H_i\Big) (\rho_S^*-\tilde{\rho}_S^*) \bigg\}\\
    &=& \tr\bigg\{ \mathcal{V}_S (\rho_S^*-\tilde{\rho}_S^*) \bigg\}.
\end{IEEEeqnarray}
But the energy of the internal sites, $i = 2, \ldots, N-1$, do not change during the heat stroke. As a consequence, 
\begin{equation}\label{LC_prop_zero_middle}
    \tr\Big\{ H_i(\rho_S^*-\tilde{\rho}_S^* )\Big\} = 0,
    \qquad i = 2,\ldots, N-1.
\end{equation}
Thus, when the system reaches the limit cycle, the energies of all internal sites no longer change, neither in the heat nor the work strokes. 
The thermodynamic output is therefore completely determined by the changes in the internal energies of the boundary sites. This is a rather peculiar feature.

\subsection{Connection to other frameworks}

In this section we discuss the connections between our framework and two other scenarios that are frequently studied in the literature.
First, our framework can be viewed as a generalization of the SWAP engine introduced in~\cite{Allahverdyan2010,Campisi2014,Uzdin2014,Campisi2015,Guarnieri2019}.  
The engine consists of two non-resonant qubits.
The heat stroke is exactly as described above, except that one assumes full thermalization; that is, the map~\eqref{eq:mapQ} thermalizes 1 to $T_C$ and 2 to $T_H$.
The work stroke is also defined in a similar way, but the unitary is now taken to be a full SWAP between the two qubits; i.e, $U_\text{SWAP} = \frac{1}{2} (1 + \sigma_x^1 \sigma_x^2 + \sigma_y^1 \sigma_y^2 + \sigma_z^1 \sigma_z^2)$, where $\sigma_\alpha^i$ are Pauli matrices of the two qubits. 

Compared to the SWAP engine~\cite{Allahverdyan2010,Campisi2014,Uzdin2014,Campisi2015,Guarnieri2019}, our scenario encompasses arbitrary Hilbert spaces for systems and ancillas, as well as arbitrary unitaries.
The use of more general Hilbert spaces allows one to explore quantum chains made up of generic $d$-level sites, as well as more exotic chain geometries. 
And the use of arbitrary unitaries allows one to consider only partial thermalization and therefore study finite-time engines and transient effects. 

Next, we compare our framework to a continuous-time scenario. 
If the duration of the heat and work strokes are sufficiently small, one may in principle move to a continuous time description by defining $\rho_S(t) = (\rho_S^{n+1} - \rho_S^n)/\tau$, where $\tau = \tau_q + \tau_w$ and $t = n\tau$. 
{\color{black}Provided the changes within the strokes are small}, for small $\tau$ the evolution of $\rho_S(t)$  {\color{black}will in general} become smooth, described in terms of a master equation~\cite{Strasberg2017}. {\color{black}A detailed comparison, which also includes 4-stroke engines, was done in Ref.~\cite{PhysRevX.5.031044}. 
}
In fact, a calculation identical to the one performed in~\cite{Barra2015,DeChiara2018,Pereira2018}, shows that the continuous-time limit of our two-stroke engine is the so-called Local Master Equation (LME) (also called boundary driven master equation), where Lindblad jump operators act only on the end-sites. 



The physical interpretation of this continuous-time limit is identical to that used in classical thermodynamics. 
A car engine, for instance, is stroke based. 
However, each cycle lasts for a very short period of time so that, in a coarse-grained time scale, one can view it as operating in continuous-time. 
Similarly, LMEs can be viewed as the continuous-time limit of our two-stroke engine.

\subsection{\label{sec:universalotto} Universal Otto efficiency}

A special feature of the SWAP engine~\cite{Allahverdyan2010,Campisi2014,Uzdin2014,Campisi2015} is that, despite having only two strokes, its efficiency is always given by the Otto efficiency. 
It turns out that there is a broader class of problems for which this is also true. 
In fact, as we now show, this will be the case whenever the internal interaction $\mathcal{V}_S$ has the form 
\begin{equation}\label{VS_Otto}
    \mathcal{V}_S = \sum\limits_{i=1}^{N-1} g_{i,i+1} \big(L_i^\dagger L_{i+1} + L_{i+1}^\dagger L_i\big),
\end{equation}
where the $L_i$ are eigenoperators of each site Hamiltonian $H_i$. That is, $[H_i, L_i] = - \omega_i L_i$. 
The transition frequency $\omega_i$ may in general be different from one site to another.
However, it is necessary for~\eqref{VS_Otto} to contain only one jump operator for each site. Mixing multiple jump operators doesn't work. 
We also notice that the local Hamiltonians $H_i$ can still be absolutely general, each with arbitrary dimensions and internal structures. 

To prove this claim, we focus on the work stroke. 
Using Heisenberg's equation, the evolution of each local site Hamiltonian will be given by 
\[
    \frac{d\langle H_i \rangle}{d t} = i \omega_i g_{i-1,i}\langle L_{i-1}^\dagger L_i - L_i^\dagger L_{i-1} \rangle - i \omega_i g_{i,i+1} \langle L_i^\dagger L_{i+1} - L_{i+1}^\dagger L_i \rangle.  
\]
Integrating over the duration $\tau_w$ of the work stroke, we find that 
\begin{equation}
    \tr\big\{ H_i (\rho_S^* - \tilde{\rho}_S^*) \big\} = \omega_i (J_{i-1,i} - J_{i,i+1}), 
\end{equation}
where
\[
    J_{i,i+1} = g_{i,i+1} \int\limits_0^{\tau_w} dt \langle L_i^\dagger L_{i+1} - L_{i+1}^\dagger L_i \rangle.
\]
For simplicity, we assumed the system was already in the limit-cycle. 
This result holds for all internal sites $i= 2, \ldots, N-1$. 
It can also hold for the boundaries, provided we define $J_{0,1} = J_{N,N+1} = 0$.

Because of the limit-cycle property~\eqref{LC_prop_zero_middle}, however, one must  have 
\begin{equation}
    J_{1,2} = J_{2,3} = \ldots = J_{N-1,N}.
\end{equation}
As a consequence, using the definitions of $\mathcal{Q}_C^*$ and $\mathcal{Q}_H^*$ in Eqs.~\eqref{QC_limit} and \eqref{QH_limit}, one finds that 
\begin{equation}\label{otto_heat_relation}
    \mathcal{Q}_C^* = \omega_1 J_{1,2} = \omega_1 J_{N,N-1} = - \frac{\omega_1}{\omega_N} \mathcal{Q}_H^*. 
\end{equation}
On the other hand, if there was no work, from Eqs.~\eqref{W_limit} and \eqref{LC_prop_zero_middle}
it is clear to note that $\mathcal{Q}_C = - \mathcal{Q}_H$. Considering nonzero work, Eq.~\eqref{otto_heat_relation} establishes a direct relation between the two heats in the limit-cycle. 
Because of the 1st law, Eq.~\eqref{1st_law_LC}, this also fixes $\mathcal{W}^*$ in terms of $\mathcal{Q}_H^*$. 

The efficiency is defined as 
\begin{equation}\label{efficiency}
    \eta = \frac{\mathcal{W}^*}{\mathcal{Q}_H^*} = 1 + \frac{\mathcal{Q}_C^*}{\mathcal{Q}_H^*}, 
\end{equation}
where we also used the 1st law~\eqref{1st_law_LC}. 
Substituting~\eqref{otto_heat_relation} then finally leads to
\begin{equation}\label{otto_efficiency}
    \eta = 1 - \frac{\omega_1}{\omega_N}, 
\end{equation}
which is the Otto efficiency~\cite{Callen}.
The engine's efficiency is therefore completely determined by the transition frequencies of the first and last sites. 
Note that these frequencies are established by the jump operators $L_i$ in Eq.~\eqref{VS_Otto}: the local Hamiltonians $H_i$ will in general have several transition frequencies. But the interaction $\mathcal{V}_S$ in~\eqref{VS_Otto} selects a specific $\omega_i$ for each site. 
We also call attention to the fact that~\eqref{otto_efficiency} is independent of the cycle duration $\tau$.
As a consequence, one may tune $\tau$ to optimize the output power, without having to bother about a decrease in efficiency. 

Eq.~\eqref{otto_heat_relation} also has an important consequence for the 2nd law. 
Substituting it in Eq.~\eqref{Sigma_LC}, one finds that 
\begin{equation}\label{Sigma_Otto}
    \Sigma^* = \bigg(\frac{\omega_1}{T_C} - \frac{\omega_N}{T_H} \bigg) \frac{\mathcal{Q}_H^*}{\omega_N}.
\end{equation}
Since $\Sigma^* \geq 0$ by construction, it follows from this result that $\mathcal{Q}_H^*$ must have the same sign as the pre-factor. 
Thus, what determines the direction of heat flow is not  the  gradient of temperature, but the ``gradient of $\omega/T$''. 
That is, the difference between $\omega_1/T_C$ and $\omega_N/T_H$. 
This is so because there is work involved, so that the standard Clausius statement, saying that heat must flow from hot to cold, does not apply (since it assumes there is no work involved). 
This result generalizes a discussion in~\cite{DeChiara2018} about possible violations of the 2nd law in LMEs~\cite{Levy2014}.
Refs.~\cite{DeChiara2018,Levy2014} dealt with  bosonic chains (fermionic chains are mathematically equivalent). In that case, what mattered for the heat flow direction was the difference in the Bose-Einstein (Fermi-Dirac) occupations. 
Eq.~\eqref{Sigma_Otto} shows that this is more general.
All it requires is an eigenoperator-type interaction of the form~\eqref{VS_Otto}. 
{\color{black} As an interesting sanity check, we may verify what happens when the Otto efficiency coincides with the Carnot efficiency. That is, when the frequencies are chosen so that $\omega_1/\omega_N = T_C/T_H$. In this case we see from Eq.~\eqref {Sigma_Otto} that $\Sigma^* = 0 $, which agrees with the idea of the Carnot cycle being reversible. 
In the SWAP engine, the output power is also identically zero in this limit, so that even though the engine operates reversibly, nothing is extracted from it. 
It is unclear to us whether a similar result should also hold for all 2-stroke engines encompassed in our framework.
}

%
%
\section{\label{sec:applications}Applications and examples}
%
%

We now illustrate our framework by considering two examples, one which can be solved analytically and another which must be handled numerically. 

%
%
\subsection{\label{sec:analytic}Analytical solution of a partial SWAP engine}
%
%

We begin by considering a system composed of two non-resonant qubits, each with local Hamiltonian  $H_{i} = \omega_i \sigma_z^i/2$. 
The ancillas for the two baths are taken to be resonant with their respective qubits. 
That is $H_C = \omega_C \sigma_z^C/2$ and $H_H = \omega_H \sigma_z^H/2$ with $\omega_C = \omega_1$ and $\omega_H = 
\omega_2$.
The initial thermal states $\rho_x$ of the two baths are thus characterized only by the Fermi-Dirac population, $f_x = (e^{\beta_x\omega_x} + 1)^{-1}$.
We take all interactions to be of the form [c.f. Eqs.~\eqref{VC_eigenop} or \eqref{VS_Otto}]
\begin{equation}\label{partial_swap_general_interaction}
\vartheta_{\mu,\nu} = g_{\mu,\nu} (\sigma_+^\mu \sigma_-^\nu + \sigma_-^\mu \sigma_+^\nu).
\end{equation}
By this we mean the internal system interaction $\mathcal{V}_S = \vartheta_{12}$ as well as the system-bath interactions $V_C = \vartheta_{1,C}$ and $V_H = \vartheta_{2,H}$.
The conditions that $1C$ and $2H$ are resonant then ensures
that there is no on/off work during the heat stroke. 
Moreover, the fact that 1 and 2 are \emph{not} resonant is precisely the source of work during the work stroke. 

Instead of working with the full map~\eqref{stroke2}, it turns out that in this case one can write down a closed system of equations for only a handful of observables for the system qubits $1$ and $2$.
We define the $c$-number variables $Z_i^n = \langle \sigma_z^i \rangle_n$ ($i = 1,2$), as well as the correlations $S^n = \langle \sigma_+^1 \sigma_-^2 + \sigma_-^1 \sigma_+^2 \rangle_n$ and 
$A^n = i\langle \sigma_+^1 \sigma_-^2 - \sigma_-^1 \sigma_+^2 \rangle_n$ {\color{black}[where $\langle \ldots \rangle_n = \tr(\ldots \rho_S^n)$]}.
From these variables, the heats are computed as $\mathcal{Q}_x^n = \omega_x(\tilde{Z}_x^n-Z_x^n)/2$ while the work is $\mathcal{W}^n =-\sum_{i=1,2} \omega_i(Z_i^{n+1} - \tilde{Z}_i^n)/2$. 

Using the map~\eqref{eq:mapQ}, a straightforward calculation shows that during the heat stroke these variables will evolve according to 
\begin{IEEEeqnarray}{rCl}
\label{diffEq_heat_Zi}
    \tilde{Z}_i^n &=& (1-\lambda) Z_i^n + \lambda Z_i^\text{th}, \\[0.2cm]
\label{diffEq_heat_S}    
    \tilde{S}^n &=& (1-\lambda) \big[\sqrt{p} S^n + \sqrt{1-p}A^n \big], \\[0.2cm]
\label{diffEq_heat_A}    
    \tilde{A}^n &=& (1-\lambda) \big[ \sqrt{p} A^n - \sqrt{1-p} S^n\big],
\end{IEEEeqnarray}
where $Z_i^\text{th} = (2 f_i-1)$ is the equilibrium spin component of each qubit in the temperature of its respective bath.
We also defined the parameters 
$p=\cos^2[(\omega_1-\omega_2)\tau_q]$ and 
$\lambda = (1 -\cos(2 g \tau_q))/2$, where {\color{black} $g = g_{CH} = g_{C,1} = g_{H,2}$} is the interaction parameter for the system-bath interactions [Eq.~\eqref{partial_swap_general_interaction}], which we assume are the same for both.

These equations help to clarify the role of different parameters, as well as the relevant time scales. 
The system-bath interactions $1C$ and $2H$ are nothing but partial SWAPs with strength $\lambda$, with $\lambda = 1$ meaning full thermalization (as is clear from Eq.~\eqref{diffEq_heat_Zi}). 
The parameter $p$, on the other hand, represents a ``transfer'' from $S$ to $A$, which is associated to the mismatch $\omega_1 - \omega_2$ between the two qubits and is independent of $g_{CH}$. 
Notice also that since the system qubits do not interact during the heat stroke, if initially $S^n = A^n = 0$, then the same will be true of $\tilde{S}^n$ and $\tilde{A}^n$;
that is to say, the heat stroke cannot create correlations between the two qubits, only destroy them. 

Similarly, during the work stroke defined by the map~\eqref{eq:mapw}, the variables $(Z_1,Z_2,S,A)$ are found to evolve according to
\begin{IEEEeqnarray}{rCl}
\label{diffEq_work_Z1}
    Z_1^{n+1} &=& (1-\eta) \tilde{Z}_1^n + \eta \tilde{Z}_2^n + 2 \eta \tan(\theta) \tilde{S}^n - 2\xi \tilde{A}^n, \\[0.2cm]
\label{diffEq_work_Z2}    
    Z_2^{n+1} &=& (1-\eta) \tilde{Z}_2^n + \eta \tilde{Z}_1^n - 2 \eta \tan(\theta) \tilde{S}^n + 2 \xi \tilde{A}^n, \\[0.2cm]
\label{diffEq_work_S}    
    S^{n+1} &=& \eta \tan(\theta) (\tilde{Z}_1^n - \tilde{Z}_2^n) + (1-2\eta\tan^2\theta) \tilde{S}^n + 2 \xi \tan(\theta) \tilde{A}^n, 
    \IEEEeqnarraynumspace
    \\[0.2cm]
\label{diffEq_work_A}    
    A^{n+1} &=& \xi(\tilde{Z}_1^n - \tilde{Z}_2^n) - 2\xi \tan(\theta) \tilde{S}^n + (1- 2 \eta \sec^2\theta) \tilde{A}^n,
\end{IEEEeqnarray}
where we introduced the auxiliary parameters 
$\eta=(2g^2/\omega_r^2)[1-\text{cos}(\omega_r \tau_w)]$ {\color{black} (not the be confused with the efficiency)}, 
$\xi = (g/\omega_r)\sin(\omega_r \tau_w)$ and $\tan(\theta) = (\omega_1- \omega_2)/2g$, with $\omega_r:=\sqrt{4g^2+(\omega_1-\omega_2)^2}$ being the Rabi frequency. 
These parameters are related according to $\xi^2 = \eta(1- \eta \sec^2\theta)$.

The parameter $\eta$ plays a similar role to $\lambda$ in Eqs.~\eqref{diffEq_heat_Zi}-\eqref{diffEq_heat_A}, quantifying the strength of the internal system coupling. 
Unlike $\lambda$, however, the parameter $\eta$ can never implement a full SWAP; that is, one can never have $\eta = 1$. In fact, $\eta < \cos^2\theta$.
This occurs because the two qubits are not resonant; if they were, then we would have $\omega_r = 2g$ and the limit $\eta \to 1$ would be reachable.
The parameter $\theta$ plays a similar role to $p$, in the sense that it is related to the detuning between the two qubits, and vanishes if they are resonant. Unlike $p$, however,  $\theta$ is independent of the interaction time.

\begin{figure}[h!]
\centering
    \includegraphics[scale=0.42]{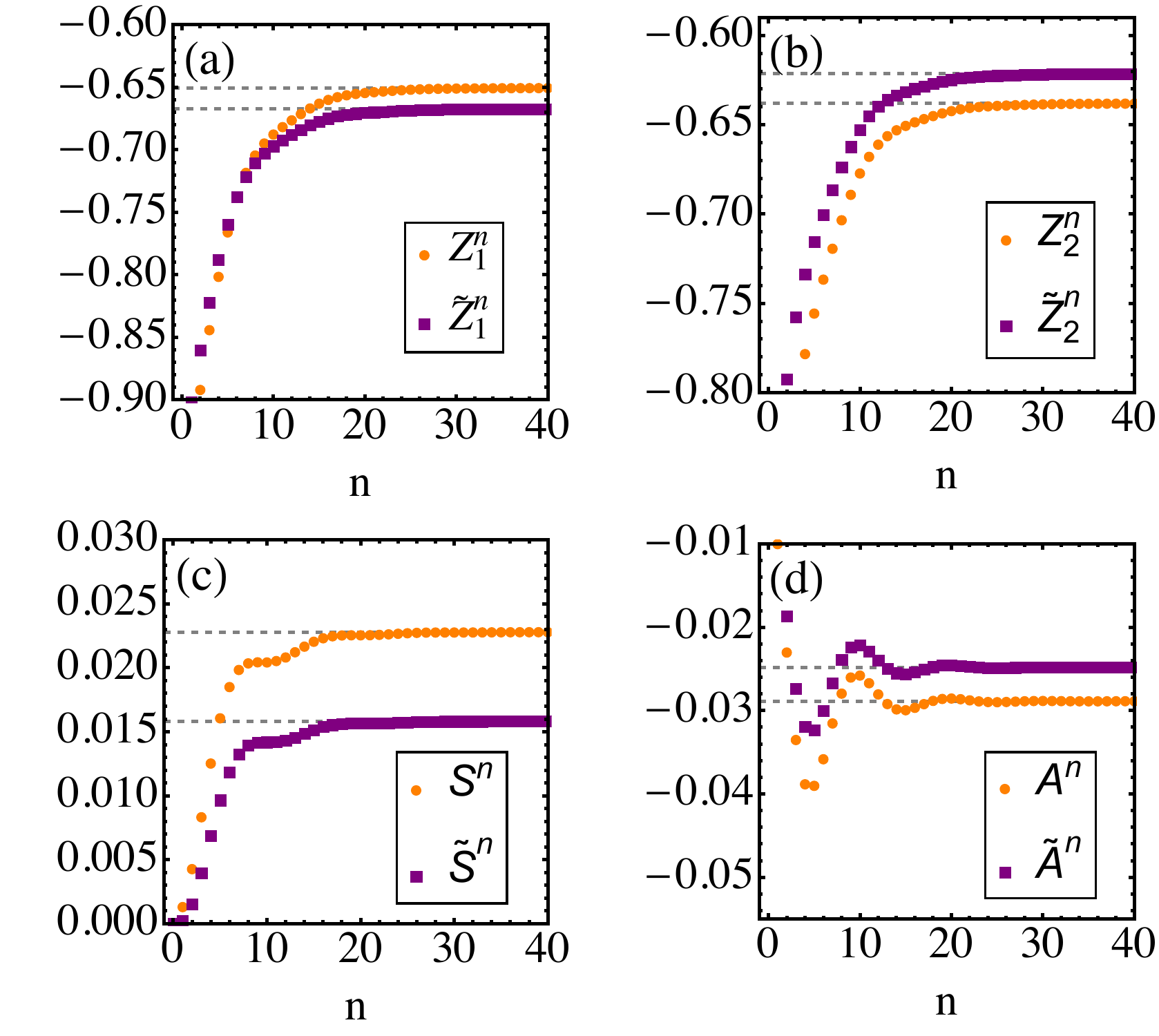}
    \caption{(Color online) Values of (a) $Z_1^n$ (``yellow circle'') , $\tilde{Z}_1^n$ (``purple square''), (b) $Z_2^n$ (``yellow circle''), $\tilde{Z}_2^n$ (``purple square'') (c) $S^n$, (``yellow circle''), $\tilde{S}^n$ (``purple square'') and (d) $A^n$ (``yellow circle''), $\tilde{A}^n$ (``purple square''), as functions of the number of cycles $n$. The dashed gray lines are the stationary values of the c-number variables. These plots were obtained for $\lambda = 0.2$, $p = 0.99$, $T_C = 0.4$, $T_H = 0.8$, $\omega_1 = 0.75$, $\omega_2 =1.0$ and $g = 0.3$. The interaction times were fixed at $\tau_q = \tau_w = 1$.
    The initial state is assumed to be both qubits in the ground-state.
    }
    \label{fig:cnumbervariables}
\end{figure}

The work stroke, as can be seen, involves not only a partial SWAP in the local populations $Z_1$ and $Z_2$, but also mixes the populations and correlations.  
This mixing is related in a not so intuitive way to both parameters $\xi$ and $\theta$. 
In fact, even in the resonant case, where $\theta = 0$, this mixing still remains because $\xi$ continues to be non-zero.

The two systems of equations, \eqref{diffEq_heat_Zi}-\eqref{diffEq_heat_A} and 
\eqref{diffEq_work_Z1}-\eqref{diffEq_work_A} form a simple set of \emph{vector difference equations} for the vector $\bm{x}_n = (Z_1^n,Z_2^n,S^n,A^n)$, which can be written as 
\begin{IEEEeqnarray}{rCl}
\label{diff_eq_1}
    \tilde{\bm{x}}_n &=& J\bm{x}_n + S, \\[0.2cm]
    \bm{x}_{n+1} &=& D \tilde{\bm{x}}_n = DJ \bm{x}_n + DS,
\label{diff_eq_2}                 
\end{IEEEeqnarray}
where the $4\times 4$ matrices $J$ and $D$, as well as the vector $S$, can be readily read from Eqs.~\eqref{diffEq_heat_Zi}-\eqref{diffEq_work_A}.
For instance, the matrix $S$ is associated only with the terms $\lambda S_i^\text{th}$ in Eq.~\eqref{diffEq_heat_Zi} and therefore reads $S = \lambda\, \text{diag}(2f_C-1,2f_H-1,0,0)$. 

The general solution of this type of difference equation reads
\begin{equation}\label{xn_sol}
    \bm{x}_n = (DJ)^n \bm{x}_0 + \sum\limits_{r=0}^{n-1} (DJ)^{n-r-1} (DS). 
\end{equation}
We therefore see that the bulk of the dynamics is governed by the matrix $DJ$. 
In addition, one may also  determine the steady-state by setting $\bm{x}_n = \bm{x}_{n+1} = \bm{x}^*$ in Eq.~\eqref{diff_eq_2}. 
As a result one finds
\begin{equation}
    \bm{x}^* = (I_4 - DJ)^{-1} DS.
\end{equation}
The full expression is somewhat cumbersome, but can nonetheless be computed analytically using e.g. \emph{Mathematica}.



\begin{figure}[h!]
    \centering
    \includegraphics[width=0.45\textwidth]{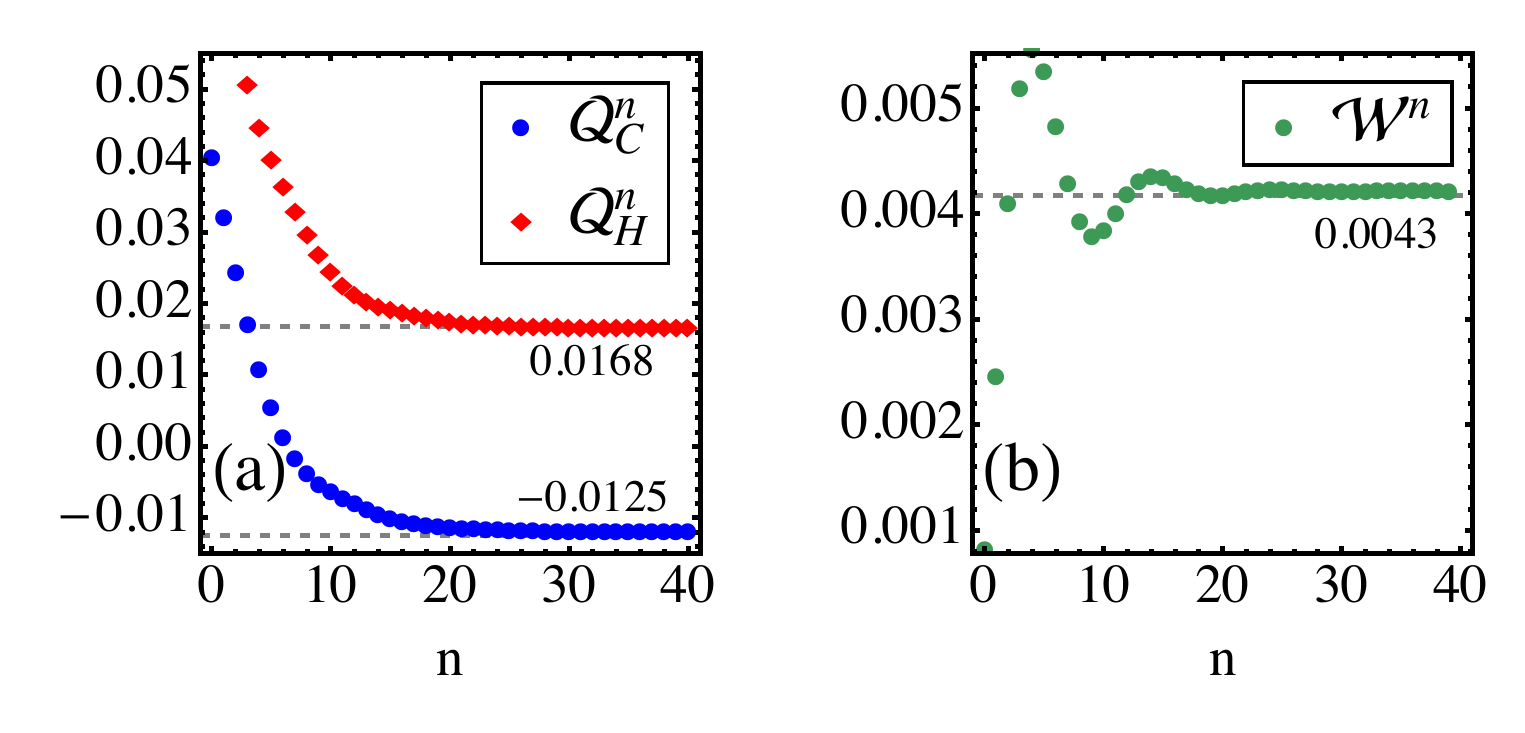}
    \caption{(Color online) Plots of (a) $\mathcal{Q}_{C}^n$ (blue circles), $\mathcal{Q}_{H}^n$ (red diamonds) and (c) $\mathcal{W}^n$ (green circles) with respect to the number of cycles $n$.  The parameters are the same as in Fig.~\ref{fig:cnumbervariables}. The numbers in the plot indicate the steady-state values for this specific configuration.
    }
    \label{fig:heatwork}
\end{figure}

An illustrative example of the evolution of the $\bm{x}_n$ and $\tilde{\bm{x}}_n$ is shown in Fig.(\ref{fig:cnumbervariables}). These plots clearly exhibit the convergence towards the steady state, where the system keeps bouncing back and forth between $\bm{x}^*$ and $\tilde{\bm{x}}^*$, similarly to a piston going up and down.
From the entries of $\bm{x}_n$ and $\tilde{\bm{x}}_n$, one readily computes the relevant thermodynamic variables $\mathcal{Q}_x = \omega_x(\tilde{Z}_x^n-Z_x^n)/2$ and $\mathcal{W} = - \sum_{i=1,2} \omega_i(Z_i^{n+1} - \tilde{Z}_i^n)/2$. The results are shown in  Fig.~\ref{fig:heatwork}.
The thermodynamic variables tend to non-zero values, as indicated by the numbers in the figure. Since the work is positive (extraction), this corresponds to a heat engine configuration.
We see that, initially, the work is very small, while the heat losses are significant. This, of course, depends on the initial conditions. But they reflect well the typical adaptation of the heat engine towards the limit-cycle operation. 
This therefore serves to illustrate that thermodynamically relevant quantities, such as the net output work in the limit-cycle, may behave in significantly different ways, in the transient and the limit cycle. 

Next we address the \emph{relaxation time}, which is associated with the number of cycles that the engine must run before reaching the limit-cycle. 
This is related to the eigenvalues of the matrix $DJ$, which is the basic matrix governing the dynamics of $\bm{x}_n$  in Eq.~\eqref{xn_sol}.
The dynamics depends on powers of $DJ$. 
Stability thus requires that its eigenvalues lie within the unit circle. 
Small eigenvalues are quickly suppressed when taking it to the power $n$. 
Thus, the longest relaxation time of the system will be described by the largest eigenvalue (in magnitude) of $DJ$:
\begin{equation}\label{mu}
    \mu := \max | \text{eigs}(DJ)|.
\end{equation}
The closer $\mu$ is to unity, the longer the system takes to relax to the limit-cycle. 

From Eqs.~\eqref{diffEq_heat_Zi}-\eqref{diffEq_heat_A}, one notices that all entries in the matrix $J$ depend equally on $1-\lambda$.
Hence, $\mu \propto 1-\lambda$. 
It turns out, however, that this is an exact equality. As may be verified from the clumsy, but exact, formulas for $D$ and $J$, the matrix $DJ/(1-\lambda)$ has an eigenvalue $1$. And so, by stability, all others must necessarily have magnitude below 1. 
Hence, we conclude that the longest relaxation time~\eqref{mu} is exactly $\mu = 1- \lambda$. 
That is, the relaxation is fully dictated by the heat stroke, as one might intuitively expect.


\begin{figure}
    \centering
    \includegraphics[scale=0.355]{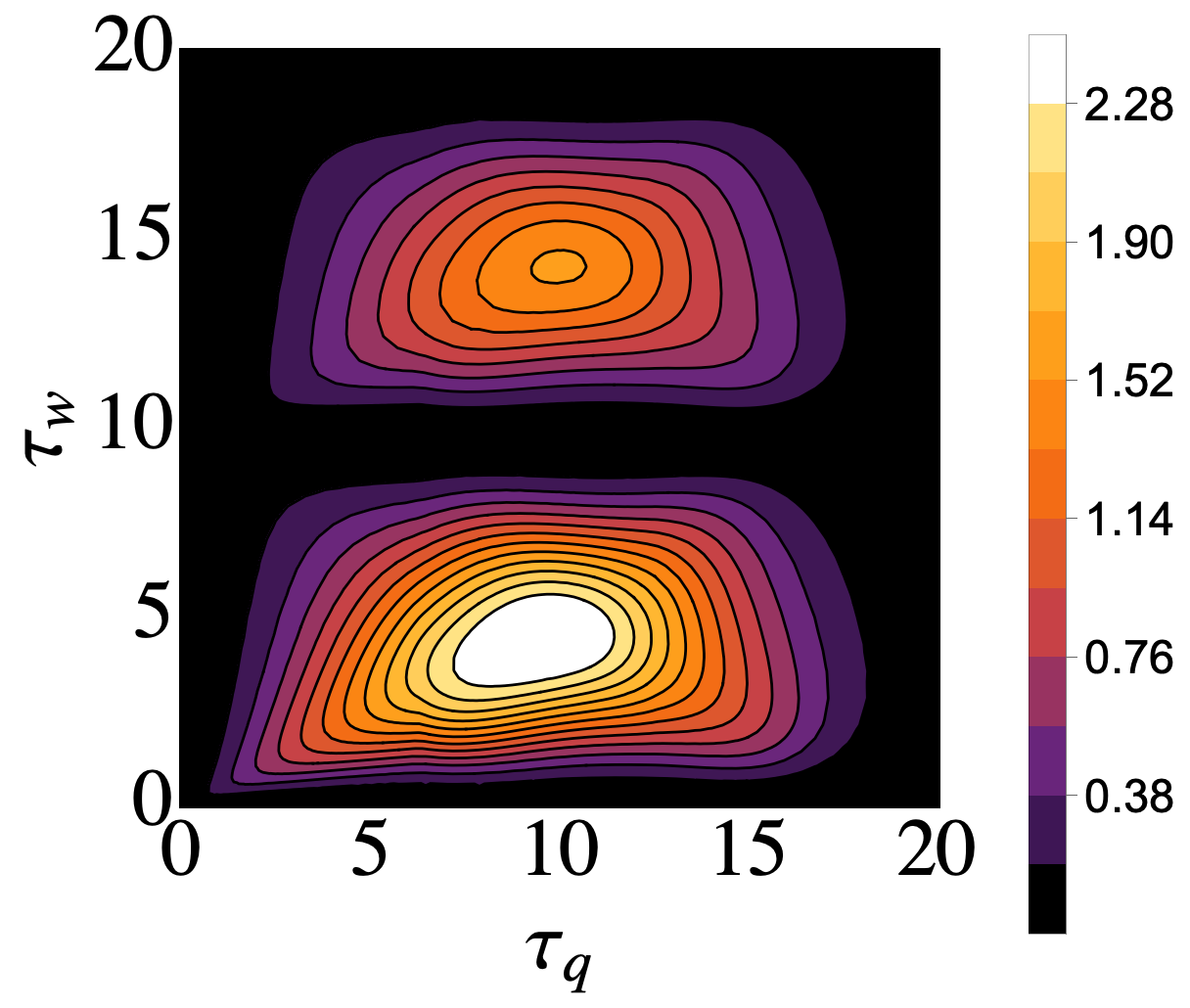}
    \caption{Optimization of the output power. The figure corresponds to a plot of Eq.~\eqref{Power} (scaled by $10^3$) as a function of the interaction times $\tau_q$ and $\tau_w$, of the heat and work strokes. The parameters are similar to those of Fig.~\ref{fig:cnumbervariables}; namely $T_C = 0.4$, $T_H = 0.8$, $\omega_1 = 0.75$, $\omega_2 = 1.$, $g = g_{CH} = 0.3$. 
    }
    \label{fig:optimization}
\end{figure}

Finally, we turn to the output power, 
\begin{equation}\label{Power}
    \mathcal{P}^* = \frac{\mathcal{W}^*}{\tau_q+\tau_w}.
\end{equation}
Our system falls under the category of Sec.~\ref{sec:universalotto} and therefore its efficiency is always given by the Otto formula~\eqref{otto_efficiency}. 
The output power, however, is not fixed but depends sensibly on all parameters. 
Using the difference equations approach described above, it is possible to write down an explicit, and not too cumbersome, formula for $\mathcal{W}^*$:
\begin{widetext}
\begin{equation}\label{W2qubits}
    \mathcal{W}^* = \frac{2 \eta (2-\lambda) \lambda(f_C-f_H) (\omega_1-\omega_2)}{\lambda^2 + 2 (1+\eta)(1-\lambda) - 2 \sqrt{p} (1-\lambda)\big[1-\eta (\tan^2\theta+\sec^2\theta)\big] + 4 \sqrt{1-p} (1-\lambda) \xi \tan\theta }.
\end{equation}
\end{widetext}
The output power depends on the stroke times $\tau_q$ and $\tau_w$, which are also contained implicitly in the parameters $\lambda$, $p$ and $\eta$ above. 

Eq.~\eqref{W2qubits} provides a clear illustration of the physics behind each parameter involved in the model. 
First, we see that $\mathcal{W}^*$ depends on $(f_C-f_H)(\omega_1-\omega_2)$. The Fermi function $(e^x+1)^{-1}$ is monotonically \emph{decreasing} in $x$. Thus, if we fix $T_C < T_H$, the output work will be positive when $f_c < f_H$ and $\omega_1 < \omega_2$, which establishes the operating interval for the machine to function as an engine:
\begin{equation}
    \frac{T_C}{T_H} \leqslant \frac{\omega_1}{\omega_2} \leqslant 1. 
\end{equation}

It is also interesting to analyze some particular cases of~\eqref{W2qubits}. 
First, for full thermalization ($\lambda \to 1$), it reduces to 
\[
\mathcal{W}^* = 2 \eta(f_C- f_H) (\omega_1-\omega_2). 
\]
This is close to the result of the SWAP engine~\cite{Allahverdyan2010,Campisi2014,Uzdin2014,Campisi2015,Guarnieri2019}. The difference is that the SWAP engine assumes a unitary stroke implementing a full SWAP, so that $\eta = 1$. 
In our case, we are assuming $\eta$ stems from a microscopic interaction so, as discussed, this limit cannot be reached. 
Similar infinitesimal expansions can also be performed for the other parameters. In particular, the structure of the numerator in Eq.~\eqref{W2qubits} shows the leading order contributions of $\lambda$, $\eta$ and $(\omega_1-\omega_2)$ will all be linear. This, of course, is physically consistent, since the work output is meant to vanish if (i) the internal system coupling vanishes, (ii) the coupling to the baths vanish and (iii) the two system qubits are resonant. 
The latter, in particular, occurs because in this limit the interaction becomes a thermal operation. 

Eq.~\eqref{W2qubits} makes it efficient to optimize the output power by tweaking the system parameters. To illustrate this, in Fig.~\ref{fig:optimization} we plot $\mathcal{P}^*$ as a function of the interaction times $\tau_q$ and $\tau_w$, of each stroke. 
The dependence on these parameters is not trivial. Part of it is oscillatory, through their dependence on the parameters $\lambda$, $p$ and $\eta$. This is clearly observed in the figure.  
But, in addition, $\mathcal{P}^*$ is also inversely proportional to $\tau_q + \tau_w$. For very large $\tau_q$ and $\tau_w$, this will cause $\mathcal{P}^*$ to decreases as one moves across the diagonal in the figure. But for intermediate values,  increases may be observed as well.


\subsection{\label{sec:Nspins}Numerical analysis of a generic XYZ spin chain }

\begin{figure}[h!]
    \centering
    \includegraphics[width=0.49\textwidth]{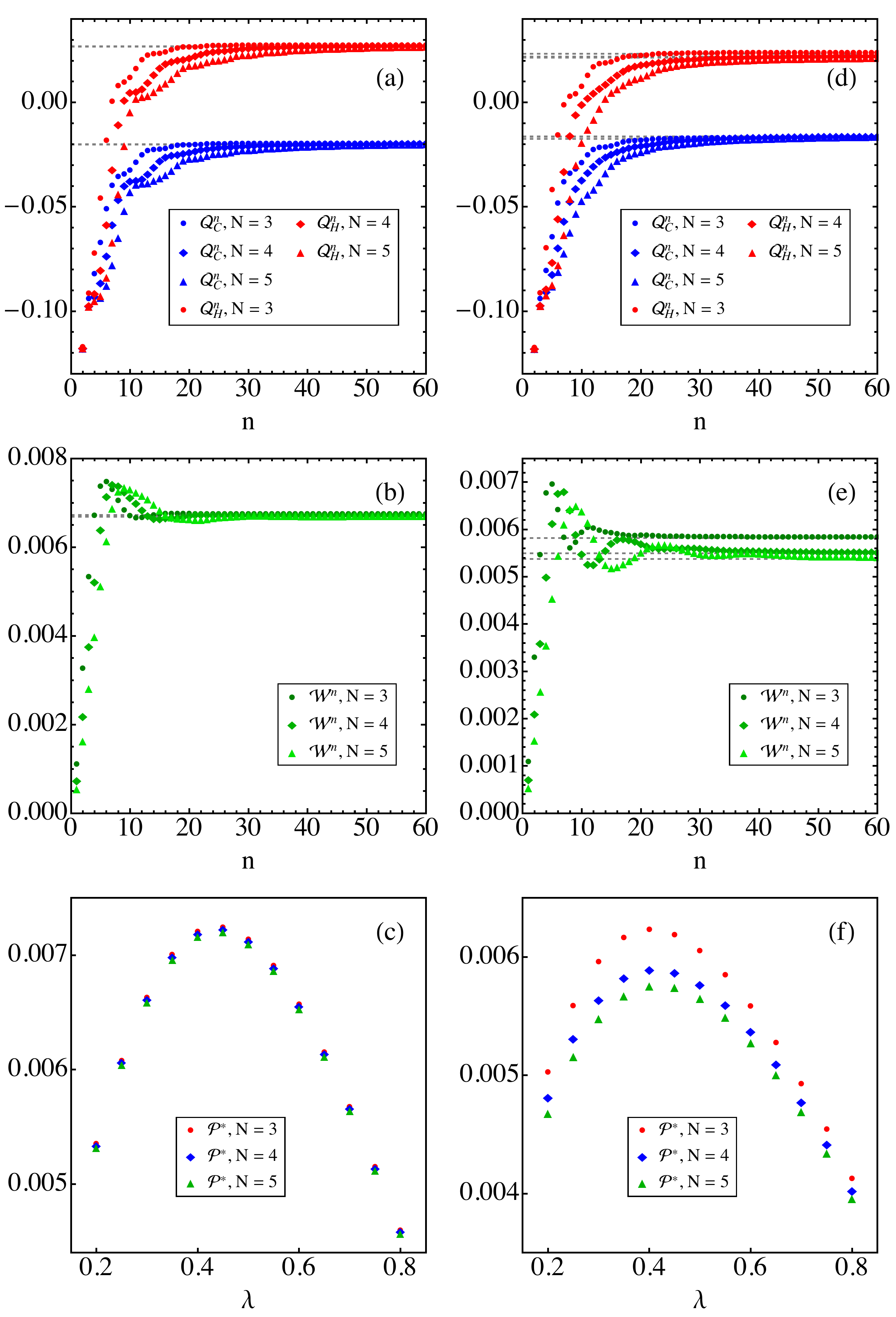}
    \caption{(Color online) Operation of a $N$-site two-stroke engine. The curves correspond to simulations for the XX (images (a)-(c)) and XXZ chain (images (d)-(f)) for different chain sizes $N$. (a),(d) presents the relaxation of the heats towards the limit cycle: $\mathcal{Q}_C^n$ blue circles ($N=3$), blue diamonds ($N=4$) and blue triangles ($N=5$); $\mathcal{Q}_H^n$: red circles ($N=3$), red diamonds ($N=4$) and red triangles ($N=5$) and (b),(e) shows the relaxation of the work $\mathcal{W}^n$: dark green circles ($N=3$), green diamonds ($N=4$) and light green triangles ($N=5$). Finally, (c),(f) presents he limit-cycle output power $\mathcal{P}^*$ as a function of $\lambda$ (which is proportional to $\tau_q$): red circles ($N=3$), blue diamonds ($N=4$) and green triangles ($N=5$). 
    The onsite potentials $\omega_i$ were chosen to interpolate linearly between $\omega_1 = 1.5$ and $\omega_N = 2.0$. Other parameters were $J_x = J_y = 0.8$, $T_C = 0.2$, $T_H = 0.8$ and $\tau_w = 0.25$. For (a)-(c) we fixed $J_z =0$ and for (d)-(f) $J_z= 0.7$. 
    }
    \label{fig:Nsites}
\end{figure}

The results for the 2-qubit engine resemble, in many aspects, the SWAP engine~\cite{Allahverdyan2010,Campisi2014,Uzdin2014,Campisi2015,Guarnieri2019}. The difference is that now the engine is operated in finite time, which therefore introduces multiple new features. 
The 2-stroke framework developed in this paper, however, is not restricted to this simple scenario. To illustrate this, we now consider an engine whose working fluid is a spin chain of $N$ sites, while the reservoirs are still single spins. 
For simplicity, we focus on linear chains, although this is not at all a restriction. 
Each spin has, as before, a local Hamiltonian  
$ H_{i} = \tfrac{1}{2} \omega_{i} \sigma_{i}^z$.
The system-bath interactions $1C$ and $NH$ are still of the form~\eqref{partial_swap_general_interaction}, with the ancillas resonant with their respective sites (so as to ensure there is no on/off work).
The internal interaction Hamiltonian however, is now taken more generally to be 
\begin{equation}
    \mathcal{V}_S = \sum_{i=1}^{N-1} \bigg\{J_x \sigma_i^x \sigma_{i+1}^x + J_y  \sigma_i^y \sigma_{i+1}^y+ J_z \sigma_i^z \sigma_{i+1}^z \bigg\}.
\end{equation}
The total Hamiltonian during the work stroke will thus be $H = \sum_i H_i + \mathcal{V}_S$. 
We focus on small chains ($N \leq 6$), for which the problem can be treated using exact diagonalization.
Moreover, we analyze two particular cases:
(i) XX model $J_x = J_y = J$, $J_z = 0$ and
(ii) XXZ chain, $J_x = J_y = J$, $J_z = J \Delta$.
The results for each case are summarized in Fig.~\ref{fig:Nsites}. 
Images (a)-(c) (left column) are for the XX and (d)-(f) (right) for the XXZ. 
Moreover, (a) and (d) show the  transient dynamics of of $\mathcal{Q}_C^n$,  $\mathcal{Q}_H^n$, while (b) and (e) show that of $\mathcal{W}^n$. 
Finally, (c) and (f) show the limit cycle output power as a function of $\lambda$ (which is equivalent to $\tau_q$).

We start by analyzing the XX model (Fig.~\ref{fig:Nsites}(a)-(c)). It can be seen that both the stationary values of heat and work are independent of the size of the chain. The extracted power in the limit cycle, on the other hand, is weakly affected by $N$ and likewise the value of $\lambda$ for which the power is maximum. These results are consistent with the fact that, by making the changes $\sigma_x \rightarrow \sigma_+ + \sigma_-$ and $\sigma_y \rightarrow -i(\sigma_+ - \sigma_-)$, the interaction Hamiltonian can be written in the same form as Eq.~\eqref{partial_swap_general_interaction} and therefore similar results to the $N=2$ case would be expected. 
{\color{black} A similarly weak dependence of the XX model in the chain size $N$  has also been observed in~\cite{DeChiara2018}, for a non-interacting bosonic chain. 
This, we believe, is related to the non-interacting nature of these models. 
For instance, it is known that the XX model presents ballistic transport, whereas the XXZ does not. 
}

Now we pass to the XXZ chain. 
In contrast with the XX model, the XXZ spin chain shows a significant variation for the heat, work and power. In Fig.~\ref{fig:Nsites}(d), we note that $\mathcal{Q}_H^*$ decreases for increasing $N$ and then, as a result, $\mathcal{W}^*$  (Fig.~\ref{fig:Nsites}(e)) also gets lower when enlarging the spin chain. A notable change in the extracted power is also observed (Fig.~\ref{fig:Nsites}(f)). The maximum value of $\mathcal{P}^*$ with respect to $\lambda$ gets lower for increasing $N$ and the $\lambda$ for which the extracted power is maximum is slightly shifted to the right.

\section{\label{sec:conclusion}Summary and Conclusions}

In this paper we have explored a framework for dealing with 2-stroke heat engines. 
The cycles are constructed by alternating between a heat stroke, which acts locally on different parts of a system, and a work stroke that couples together the different sites.
This corresponds to a generalization of the SWAP engine~\cite{Allahverdyan2010,Campisi2014,Uzdin2014,Campisi2015,Guarnieri2019}.
Our approach is based on a collisional model, which allows us to properly take into account all energy changes in the system and thus completely characterize heat, work and entropy production.  
As we show, this framework is particularly suited for capturing the finite-time dynamics of the system and the convergence toward a limit cycle. 
In particular, we establish a broad class of models which present a universal Otto efficiency. 
As an application, we study a finite-time generalization of the 2-qubit SWAP engine, as well as a spin chain of $N$ sites and different types of interactions. 
We show that the system may present a rich set of behaviours, as well as operating regimes (engine, refrigerator etc.) depending on the choices of parameters.


\section*{\label{sec:acknowledgements}Acknowledgements}

The authors thank R. Uzdin, N. Myers and S. Deffner for fruitful correspondence. 
G.T.L. acknowledges support from the São Paulo Research Foundation (FAPESP) through Grants No. 2018/12813-0, No. 2017/50304-7, and No. 2017/07973-5. O.A.D.M. acknowledges support from the Brazil's National Council for Scientific and Technological Development (CNPq) through Grant No. 135905/2019-2. G.T.L. acknowledges the hospitality of apt44, where part of this work was developed. 
O.A.D.M. expresses his gratitude to Dr. M. Hüber's group at IQOQI-Vienna, for the hospitality and fruitful discussions during a scientific visit.

\appendix
{\color{black}
\section{\label{app:SEC}Strict energy conservation and on/off work}

In this appendix we show that the strict energy conservation condition~\eqref{eq:localbalance} implies that the on/off work in Eqs.~\eqref{Wc_on_off} and \eqref{Wh_on_ff} must necessarily vanish. 
It suffices to focus on just two systems, 1 and 2, with Hamiltonians $H_1$ and $H_2$ and interacting through an operator $V$ satisfying strict energy conservation $[V,H_1+H_2]=0$.
The generalization to include both baths is straightforward. 

The fact that the global dynamics of 12 is unitary implies that 
\begin{equation*}
 \Delta H_1 + \Delta H_2 + \Delta V =0,
\end{equation*}
where $\Delta \mathcal{O} = \tr \big\{ \mathcal{O} (U \rho U^\dagger - \rho) \big\}$ is the change in operator $\mathcal{O}$ due to the unitary evolution generated by $U = e^{-i (H_1+H_2+V)t}$.

But if $[V,H_1+H_2] = 0$, it follows that $[U,H_1+H_2] = 0$, which in turn implies that  $(U^\dagger H_1 U - H_1) + (U^\dagger H_2 U - H_2) = 0$. Hence, we must also have 
\begin{equation*}
    \Delta H_1 + \Delta H_2 = 0. 
\end{equation*}
Comparing the two results we conclude that 
\begin{equation*}
    \Delta V = 0.
\end{equation*} 
But $\Delta V$ is precisely the on/off work (the energy trapped in the interaction). Whence, the on/off work vanishes for strict energy conservation, which is what we set out to prove. 
}

\bibliography{biblio}

\end{document}